\documentclass[amsmath,amssymb,aps,10pt,prl,letterpaper,bibnotes,balancelastpage,notitlepage,twocolumn,floatfix,superscriptaddress,nofootinbib,preprintnumbers]{revtex4-2}
\usepackage{graphicx}	% Standard figures 
\usepackage{ragged2e}	
\usepackage{bm}		    % Bold math
\usepackage[sort&compress]{natbib}	 	% Standard reference package
	\setcitestyle{square,numbers,comma}	% Set citation style
\usepackage[colorlinks=true,urlcolor=blue,linkcolor=black,citecolor=blue]{hyperref}	
\usepackage{xcolor}
\usepackage{fontawesome5}
\definecolor{orcidlogocol}{rgb}{0.65, 0.807, 0.223}
\newcommand{\orcid}[1]{\,\href{https://orcid.org/#1}{\textcolor{orcidlogocol}{\footnotesize\faOrcid}}\,}
\newcommand{\LL}{\mathcal{L}}
\newcommand{\eV}{\mathrm{eV}}
\newcommand{\FF}{\mathcal{F}}
\newcommand{\gB}{\ensuremath{\tilde{g}_{B}}}
\newcommand{\gL}{\ensuremath{\tilde{g}_{L}}}
\newcommand{\gBL}{\ensuremath{\tilde{g}_{B-L}}}
\newcommand{\gQ}{\ensuremath{\tilde{g}_{Q}}}
\newcommand{\gEff}{\ensuremath{\tilde{g}_{\oplus}}}
\newcommand{\gA}{\ensuremath{g_{\phi \gamma}}}
\renewcommand{\section}[1]{\phantomsection\addcontentsline{toc}{section}{#1}\textit{#1}.---\unskip\ignorespaces}
\newcommand{\tabref}[2][]{Table{#1}~\ref{#2}}		% Table reference
\newcommand{\figref}[2][]{Fig{#1}.~\ref{#2}}		% Figure reference
\renewcommand{\eqref}[2][]{Eq{#1}.~(\ref{#2})}		% Equation reference
\newcommand{\citeR}[2][]{Ref{#1}.~\cite{#2}}		% Ref. Citation
\newcommand{\lb}{\ensuremath{\left}}				% Left Brackets
\newcommand{\rb}{\ensuremath{\right}}				% Right Brackets
\begin{document}
%%%%%%%%%%%%%%%%%%%%%%%%%%%%%%%%%%%%%%%%%%%%%%%%%%%%%%%%%%%%%%%%%%%%%%%%%%%%%%%%%%%%%%%%%%	
%%%%%%%%%%%%%%%%%%%%%%%%%%%%%%%%%%%%%%%%%%%%%%%%%%%%%%%%%%%%%%%%%%%%%%%%%%%%%%%%%%%%%%%%%%	
%%%%%%%%%%%%%%%%%%%%%%%%%%%%%%%%%%%%%%%%%%%%%%%%%%%%%%%%%%%%%%%%%%%%%%%%%%%%%%%%%%%%%%%%%%	
%%%%%%%%%%%%%%%%%%%%%%%%%%%%%%%%%%%%%%%%%%%%%%%%%%%%%%%%%%%%%%%%%%%%%%%%%%%%%%%%%%%%%%%%%%

%%%%%%%%%%%%%%%%%%%%%%%%%%%%%%%%%%%
\preprint{FERMILAB-PUB-23-162-SQMS}
%%%%%%%%%%%%%%%%%%%%%%%%%%%%%%%%%%%

%%%%%%%%%%%%%%%%%%%%%%%%%%%%%%%%%%%%%%%%%%%%%%%%%%%%%%%%%%%%%%%%%%%%%%%%%%%%%%%%%%%%%%%%%%	
% Title, Author and Affiliation
\title{Measuring axion gradients with photon interferometry (MAGPI)}
\date{\today}
%%%%%%%%%%%%%%%%%%%%%%%%%%%%%%%%%%%%%%%%%%%%%%%%%%%%%%%%%%%%%%%%%%%%%%%%%%%%%%%%%%%%%%%%%%	

%%%%%%%%%%%%%%%%%%%%%%%%%%%%%%%%%%%%%%%%%%%%%%%%%%%%%%%%%%%%
\author{Michael A.~Fedderke\orcid{0000-0002-1319-1622}}
\email{mfedderke@perimeterinstitute.ca}
\thanks{author contributed equally.}
\affiliation{The William H.~Miller III Department of Physics and Astronomy, The Johns Hopkins University, Baltimore, Maryland 21218, USA}
\affiliation{Perimeter Institute for Theoretical Physics, Waterloo, Ontario, N2L 2Y5, Canada}
%%%%%%%%%%%%%%%%%%%%%%%%%%%%%%%%%%%%%%%%%%%%%%%%%%%%%%%%%%%%
\author{Jedidiah O.~Thompson\orcid{0000-0002-7342-0554}}
\email{jedidiah@stanford.edu}
\thanks{author contributed equally.}
\affiliation{Stanford Institute for Theoretical Physics, Stanford University, Stanford, California 94305, USA}
%%%%%%%%%%%%%%%%%%%%%%%%%%%%%%%%%%%%%%%%%%%%%%%%%%%%%%%%%%%%
\author{Raphael Cervantes\orcid{0000-0003-1386-1005}}
\affiliation{Fermi National Accelerator Laboratory, Batavia, Illinois 60510, USA}
%%%%%%%%%%%%%%%%%%%%%%%%%%%%%%%%%%%%%%%%%%%%%%%%%%%%%%%%%%%%
\author{Bianca Giaccone\orcid{0000-0002-7275-8465}}
\affiliation{Fermi National Accelerator Laboratory, Batavia, Illinois 60510, USA}
%%%%%%%%%%%%%%%%%%%%%%%%%%%%%%%%%%%%%%%%%%%%%%%%%%%%%%%%%%%%
\author{Roni Harnik\orcid{0000-0001-7293-7175}}
\affiliation{Theory Division, Fermi National Accelerator Laboratory, Batavia, Illinois 60510, USA}
\affiliation{Superconducting Quantum Materials and Systems Center (SQMS), Fermi National Accelerator Laboratory, Batavia, Illinois 60510, USA}
%%%%%%%%%%%%%%%%%%%%%%%%%%%%%%%%%%%%%%%%%%%%%%%%%%%%%%%%%%%%
\author{David E.~Kaplan\orcid{0000-0001-8175-4506}}
\affiliation{The William H.~Miller III Department of Physics and Astronomy, The Johns Hopkins University, Baltimore, Maryland 21218, USA}
%%%%%%%%%%%%%%%%%%%%%%%%%%%%%%%%%%%%%%%%%%%%%%%%%%%%%%%%%%%%
\author{Sam Posen\orcid{0000-0002-6499-306X}}
\affiliation{Fermi National Accelerator Laboratory, Batavia, Illinois 60510, USA}
%%%%%%%%%%%%%%%%%%%%%%%%%%%%%%%%%%%%%%%%%%%%%%%%%%%%%%%%%%%%
\author{Surjeet Rajendran\orcid{0000-0001-9915-3573}}
\affiliation{The William H.~Miller III Department of Physics and Astronomy, The Johns Hopkins University, Baltimore, Maryland 21218, USA}
%%%%%%%%%%%%%%%%%%%%%%%%%%%%%%%%%%%%%%%%%%%%%%%%%%%%%%%%%%%%

%%%%%%%%%%%%%%%%%%%%%%%%%%%%%%%%%%%%%%%%%%%%%%%%%%%%%%%%%%%%%%%%%%%%%%%%%%%%%%%%%%%%%%%%%%
% Abstract
\begin{abstract}
%%%%%%%%%%%%%%%%%%%%%%%%%%%%%%%%%%%%%%%%%%
We propose a novel search technique for axions with a $CP$-violating monopole coupling~$\gQ$ to bulk Standard Model charges $Q \in \{B,L,B-L\}$.
Gradients in the static axion field configurations sourced by matter induce achromatic circular photon birefringence via the axion--photon coupling~$\gA$.
Circularly polarized light fed into an optical or (open) radio-frequency (RF) Fabry--P{\'e}rot~(FP) cavity develops a phase shift that accumulates up to the cavity finesse: the fixed axion spatial gradient prevents a cancellation known to occur for an axion dark-matter search.
The relative phase shift between two FP cavities fed with opposite circular polarizations can be detected interferometrically. 
This time-independent signal can be modulated up to non-zero frequency by altering the cavity orientations with respect to the field gradient.
Multi-wavelength co-metrology techniques can be used to address chromatic measurement systematics and noise sources.
With Earth as the axion source, we project reach beyond current constraints on the product of couplings $\gQ \gA$ for axion masses $m_{\phi} \lesssim 10^{-5} \,\eV$.
If shot-noise-limited sensitivity can be achieved, an experiment using high-finesse RF FP cavities could reach a factor of $\sim 10^{5}$ into new parameter space for $\gQ \gA$ for masses $m_\phi \lesssim 4\times 10^{-11}\,\eV$.
%%%%%%%%%%%%%%%%%%%%%%%%%%%%%%%%%%%%%%%%%%
\end{abstract}
%%%%%%%%%%%%%%%%%%%%%%%%%%%%%%%%%%%%%%%%%%%%%%%%%%%%%%%%%%%%%%%%%%%%%%%%%%%%%%%%%%%%%%%%%%

\maketitle

%%%%%%%%%%%%%%%%%%%%%%%%%%%%%%%%%%%%%%%%%%%%%%%%%%%%%%%%%%%%%%%%%%%%%%%%%%%%%%%%%%%%%%%%%%
%%%%%%%%%%%%%%%%%%%%%%%%%%%%%%%%%%%%%%%%%%%%%%%%%%%%%%%%%%%%%%%%%%%%%%%%%%%%%%%%%%%%%%%%%%
\section{Introduction}%
\label{sec:introduction}%
%%%%%%%%%%%%%%%%%%%%%%%%%%%%%%%%%%%%%%%%%%%%%%%%%%%%%%%%%%%%%%%%%%%%%%%%%%%%%%%%%%%%%%%%%%
%%%%%%%%%%%%%%%%%%%%%%%%%%%%%%%%%%%%%%%%%%%%%%%%%%%%%%%%%%%%%%%%%%%%%%%%%%%%%%%%%%%%%%%%%%
Axions%
%%%%%%%%%%%%
\footnote{\label{ftnt:AxionsALPS}%
    In this work, the term `axion' is used to refer to either the QCD axion or axion-like particles (ALPs), as appropriate.} %
%%%%%%%%%%%%
are compelling extensions to the Standard Model~(SM) that can address the strong $CP$ problem~\cite{Peccei:1977hh,Weinberg:1977ma,Wilczek:1977pj}, behave as quintessence fields~\cite{Frieman:1995pm,Caldwell:1997ii} or dark-matter~(DM) candidates~\cite{Preskill:1982cy,Abbott:1982af,Dine:1982ah,Duffy:2009ig,Hui:2016ltb}, or simply be other light degrees of freedom expected from UV theory~\cite{Svrcek:2006yi,Arvanitaki:2009fg}. 

An axion $\phi$ generically couples to photons via a pseudoscalar coupling $\LL \supset - \gA \phi F \tilde{F}/4$, inducing%
%%%%%%%%%%%%
\footnote{The axion--photon--photon vertex naturally involves three fields. 
To see the birefringence effect, one considers the axion to be a background field acting to modify the photon propagation. 
This differs from the setup in many other axion-detection experiments (e.g., CAST~\cite{CAST:2017uph}, ADMX~\cite{ADMX:2018gho,ADMX:2019uok,ADMX:2021nhd}, etc.), where an applied background electromagnetic field allows conversion of a propagating axion to a detectable photon.}
%%%%%%%%%%%%
circular photon birefringence~\cite{Carroll:1989vb,PhysRevD.43.3789,Harari:1992ea,Carroll:1998zi}.
Oppositely handed photons propagating in a varying axion field pick up light-frequency-independent (achromatic) phase shifts $\alpha_{\pm}$ of opposite signs, which depend only on the difference $\Delta \phi$ in the axion field value at the endpoints of the photon path: $\alpha_\pm = \pm \gA \Delta \phi/2$.
A large body of literature exists on the phenomenology of this effect in cosmological~\cite{PhysRevLett.83.1506,Liu:2006uh,Ni:2007ar,Pospelov:2008gg,Finelli:2008jv,Galaverni:2009zz,2011PhRvD..84d3504C,2012PhRvD..86j3529G,Li:2013vga,Lee:2013mqa,Gubitosi:2014cua,Galaverni:2014gca,Gruppuso:2015xza,Aghanim:2016fhp,Sigl:2018fba,Fedderke:2019ajk,Agrawal:2019lkr,Gruppuso:2020kfy,Jain:2021shf,BICEPKeck:2021sbt,Jain:2022jrp,Eskilt:2022cff,Bortolami:2022whx,Diego-Palazuelos:2022dsq,SPT-3G:2022ods,Galaverni:2023zhv,POLARBEAR:2023ric}, astrophysical~\cite{Alighieri:2010eu,Galaverni:2014gca,Fujita:2018zaj,Ivanov:2018byi,Liu:2019brz,Liu:2021zlt}, and lab-based contexts~\cite{DeRocco:2018jwe,Melissinos:2008vn,Obata:2018vvr,Liu:2018icu,Martynov:2019azm,Nagano:2019rbw,Oshima:2023csb}.

In general, axions can also have scalar monopole couplings~$\gQ$ to bulk SM matter charges $Q$: e.g., to baryon number $B$ via $\LL \supset - \gB \phi \bar{N} N$, where $N$ is a nucleon field. 
Such couplings source static axion field gradients around ordinary objects, with a range $r_\phi \sim 1/m_{\phi}$ set by the axion mass $m_{\phi}$. 
This scenario, while violating both $CP$ and the axion shift symmetry, arises naturally for a QCD axion in the presence of a nonzero strong $\Theta$ angle~\cite{PhysRevD.30.130,Pospelov:1997uv}.
For axions with pseudoscalar couplings to fermion spins (the `monopole--dipole' scenario~\cite{PhysRevD.30.130,Agrawal:2022wjm}), the presence of such scalar couplings has also recently been invoked as a possible environmental explanation for the $(g-2)_{\mu}$ anomaly~\cite{Davoudiasl:2022gdg,Agrawal:2022wjm}; this scenario can also be searched for directly in other ways~\cite{Arvanitaki:2014dfa,ARIADNE:2017tdd,ARIADNE:2020wwm}.
A light axion with scalar couplings may involve a mass tuning, but this depends on the UV cutoff and other assumptions.

In this work, we assume the presence of both the scalar coupling $\gQ$ and the pseudoscalar coupling $\gA$.
Of course, independent constraints on both $\gA$ and $\gQ$ exist.
Bounds on $\gA$ arise from many phenomena~\cite{PhysRevD.37.1237,Raffelt:1996wa,Wouters:2013hua,Marsh:2017yvc,Reynolds:2019uqt,Reynes:2021bpe,Dessert:2022yqq,Jaeckel:2017tud,Hoof:2022xbe,Payez:2014xsa,Ayala:2014pea,Dolan:2022kul,AxionLimits}:
for example, astrophysical bodies could emit axions which may also convert to photons in magnetic fields. 
Axion--photon interconversion would also cause spectral distortions of sources.
The strongest bounds are $\gA \lesssim 6\times 10^{-13}\,\text{GeV}^{-1}$ for $m_\phi \lesssim 10^{-11}\,\eV$~\cite{Reynes:2021bpe}.
Monopole couplings $\gQ$ induce fifth forces stringently constrained by, e.g., tests of the weak equivalence principle~\cite{Schlamminger:2007ht,ADELBERGER2009102,Wagner:2012ui,PhysRevLett.129.121102,Touboul:2017grn,Berge:2017ovy,Shaw:2021gnp,Fedderke:2022ptm}: $\gB \lesssim 6\times 10^{-25}$ for $m_{\phi} \lesssim 1/R_{\oplus} \sim 3\times 10^{-14}\,\eV$~\cite{Shaw:2021gnp}, with bounds on $\gL$ and $\gBL$ being of similar magnitude. 

\begin{figure}[t]
    \centering
    \includegraphics[width=\columnwidth]{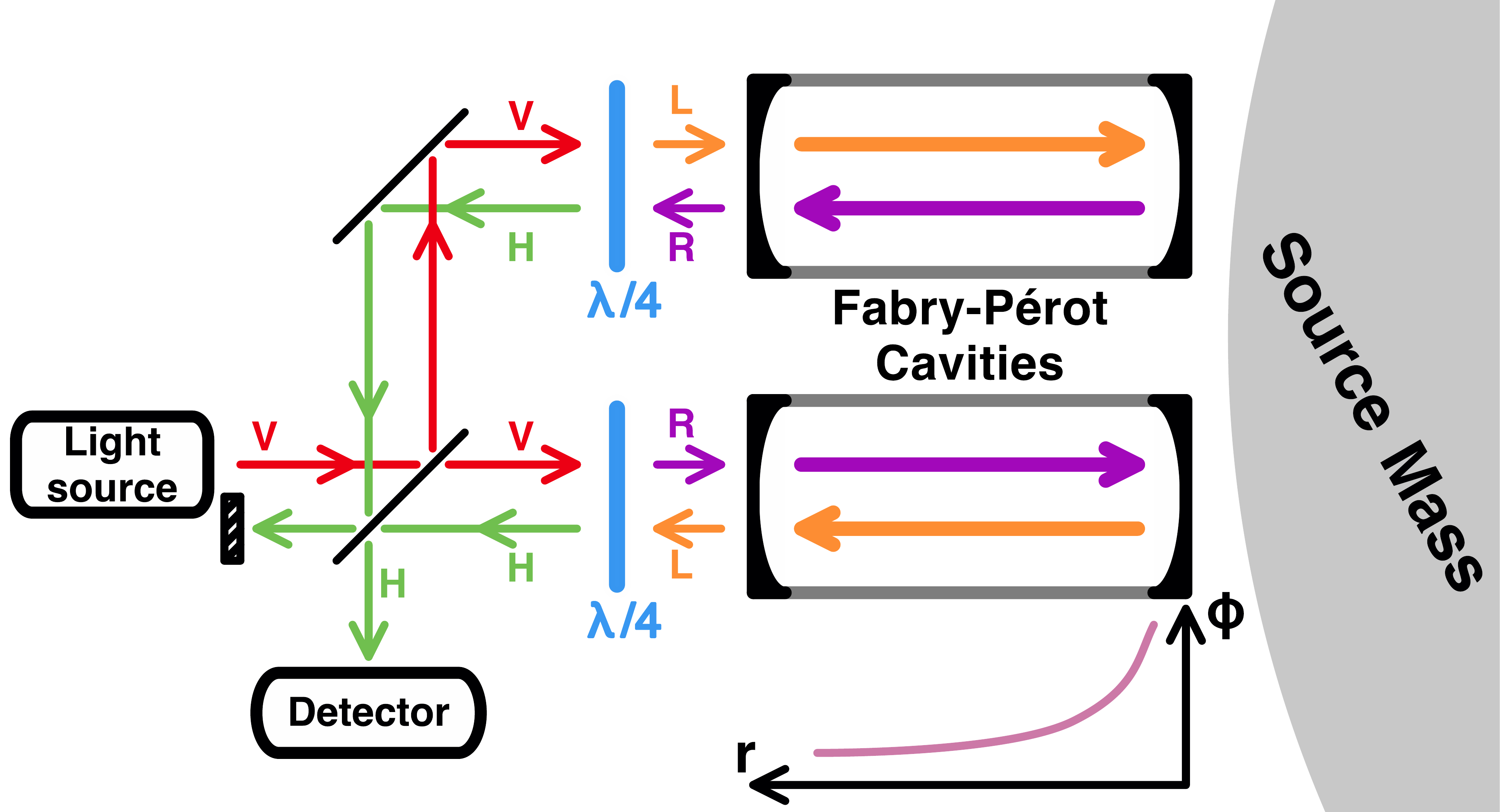}
    \caption{\label{fig:expdiagram}%
        %%%%%%%%%%%%%%%%%%%%%%%%%%%%%%%%%%
        The proposed experiment. 
        Vertically polarized~(V) light (optical or RF) is phase modulated and split into two beams, which are passed through oppositely oriented quarter-wave plates to become right- (R) and left- (L) circularly polarized, respectively. 
        The circular polarizations are fed into rigid, high-finesse FP cavities, where they accumulate a phase shift from the axion field gradient sourced by a nearby mass (e.g., Earth, or a small laboratory mass). 
        Finally, the beams pass back through the $\lambda/4$-plates to become horizontally polarized~(H) and are interfered at the beamsplitter.
        Signal detection is via a carrier--sideband beat note at the dark port.%
        %%%%%%%%%%%%%%%%%%%%%%%%%%%%%%%%%%
        }
\end{figure}

In this paper, we propose an interferometric laboratory experiment that we estimate to be capable of probing new parameter space for the product of couplings~$\gQ\gA$.
The idea is summarized in \figref{fig:expdiagram}.
A Michelson interferometer with rigid, high-finesse Fabry--P{\'e}rot (FP) cavities of length $\ell$ in its arms is set up such that those arms are fed with light of opposite circular polarization.
In the presence of a static axion field gradient sourced by nearby matter, a time-independent, achromatic phase difference develops between the two arms.
A laboratory mass could be the axion source, but the strongest accessible source is generally Earth, which for $Q=B$ gives a vertical surface field gradient $|\nabla \phi|_{\oplus} \sim 0.2\,\eV^2 \times \gB/(6\times 10^{-25})$ for $m_\phi \lesssim  1/R_{\oplus}$ (for $Q=L$ or $B-L$, $|\nabla \phi|_{\oplus}$ is smaller by a factor of $\sim 2$).
To be sensitive to this gradient, the cavities should be oriented vertically.

%%%%%%%%%%%%%%%%%%%%%%%%%%%%%%%%%%%%%%%%%%%%%%%%%%%%%%%%%%%%%%%%%%%%%%%%%%%%%%%%%%%%%%%%%%
%%%%%%%%%%%%%%%%%%%%%%%%%%%%%%%%%%%%%%%%%%%%%%%%%%%%%%%%%%%%%%%%%%%%%%%%%%%%%%%%%%%%%%%%%%

Crucially for this setup, the phase shifts in the interferometer arms accumulate linearly with cavity finesse $\FF$. 
Photon helicity flips on reflection at each end of the cavity; if uncompensated, this would cause the phase shifts for outbound and return trips in the cavity to have opposite signs, leading to a round-trip signal cancellation~\cite{DeRocco:2018jwe}.
However, that sign change is exactly compensated for by $\Delta \phi$ changing sign for trips in opposite directions in a static field gradient. 
The signal thus does not cancel, and can be boosted by working with high-finesse FP cavities; e.g., those that operate at radio frequencies (RF)~\cite{KuhrUltrahighFinesse}.

To address systematic issues, the static signal can be modulated to finite frequency by rotating the cavity orientations with respect to the gradient.
We also propose injecting multiple cavity-resonant frequencies to perform cavity `co-metrology' and break possible degeneracies with chromatic systematic effects and noise sources.

Interferometric detection of axions is not a new idea. 
A similar broadband experimental proposal for probing axion DM appears in \citeR{DeRocco:2018jwe}; see also \citeR[s]{Melissinos:2008vn,Obata:2018vvr,Liu:2018icu,Martynov:2019azm,Nagano:2019rbw,Oshima:2023csb} for other experimental designs and/or resonant approaches. 
These proposals are, however, tailored to search for the dominant, temporal component of the DM axion gradient: $|\dot{\phi}|_{\textsc{dm}} \sim 2\times 10^{-3} \,\eV^2 \sim 10^3 |\nabla \phi|_{\textsc{dm}}$.
This is smaller than the largest allowed spatial gradient from an Earth-sourced axion, and does not have an intrinsic spatial directionality.
For slowly varying DM axion fields, $m_\phi \ell \ll 1$, the latter fact leads to the round-trip signal cancellation mentioned above for this setup~\cite{DeRocco:2018jwe}; to overcome this for a DM axion, additional optical elements inside the cavity are required~\cite{DeRocco:2018jwe}.
Moreover, a DM axion experiment has a bandwidth limitation arising from the virialized nature of DM~\cite{Foster:2017hbq,Centers:2019dyn,JacksonKimball:2023jpq}. 
Such a limitation is not present for a static, sourced axion field gradient, for which noise may thus be more efficiently averaged down.

%%%%%%%%%%%%%%%%%%%%%%%%%%%%%%%%%%%%%%%%%%%%%%%%%%%%%%%%%%%%%%%%%%%%%%%%%%%%%%%%%%%%%%%%%%
%%%%%%%%%%%%%%%%%%%%%%%%%%%%%%%%%%%%%%%%%%%%%%%%%%%%%%%%%%%%%%%%%%%%%%%%%%%%%%%%%%%%%%%%%%
\section{Signal}%
\label{sec:signal}%
%%%%%%%%%%%%%%%%%%%%%%%%%%%%%%%%%%%%%%%%%%%%%%%%%%%%%%%%%%%%%%%%%%%%%%%%%%%%%%%%%%%%%%%%%%
%%%%%%%%%%%%%%%%%%%%%%%%%%%%%%%%%%%%%%%%%%%%%%%%%%%%%%%%%%%%%%%%%%%%%%%%%%%%%%%%%%%%%%%%%%
We consider the low-energy effective axion model (we work in natural units; $\hbar=c=1$)
%%%%%%
\begin{align}
\begin{split}
    \LL &\supset \LL_{\textsc{sm}} + \frac{1}{2} (\partial \phi)^2 - \frac{1}{2} m_\phi^2 \phi^2 - \frac{\gA}{4} \phi F \widetilde{F} \\
    &\quad - (\gB+\gBL) \phi \bar{N}N - (\gL - \gBL) \phi \bar{e}e \: ,
\end{split}
\end{align}
%%%%%%
where $\phi$ is the axion, $F$ ($\widetilde{F}$) is the (dual) photon field-strength tensor, $N$ is a nucleon, and $e$ is the electron.
Consider an FP cavity of length $\ell$ and finesse $\FF$ with its axis aligned to local vertical on the surface of Earth.
External laser light with wavelength $\lambda$ (wavenumber $k = 2\pi / \lambda$) is introduced to the FP cavity via the upper mirror.
We model the lower (upper) mirror to be perfectly (highly) reflecting for light incident from the interior of the cavity, and both mirrors to be lossless.

Ignoring transverse spatial gradients in the beam profile, plane right ($+$) [left ($-$)] circularly polarized light incident on the exterior face of the cavity input mirror (i.e., the upper mirror) is reflected from the cavity as left [right] circularly polarized light with a phase shift $e^{i \alpha_{+}}$ [$e^{i \alpha_{-}}$], where
%%%%%%
\begin{align}
    \tan \alpha_{\pm} &= \frac{\pi \sqrt{4\FF^2 + \pi^2} \sin( 2k\ell \pm \gA \Delta \phi )}{(2\FF^2+\pi^2) \cos( 2k\ell \pm \gA \Delta \phi ) - 2\FF^2} \: , \label{eq:phaseShiftExact}
\end{align}
%%%%%%
where $\Delta \phi \equiv \phi(\text{lower}) - \phi(\text{upper})$ is the difference in the axion field between the lower and upper mirrors.

For a cavity near longitudinal resonance, $\ell = \ell_n + \Delta \ell$ with $\ell_n = n\lambda/2 \; (n=1,2,\ldots)$ and $|\Delta \ell| \ll \ell_n$, and in the combined limits $\alpha_{\pm} \ll 1$ and $\FF\gg 1$, the phase shifts are%
%%%%%%%%%%%%
\footnote{\label{ftnt:factorOf2}%
    This result for the phase shift for the light \emph{reflected} from the cavity is a factor of 2 larger than the phase shift that would be obtained for the light \emph{transmitted} through the far mirror in a setup that employs FP cavity end mirrors with equal reflection coefficients; cf.~the PVLAS experiment~\cite{Ejlli:2020yhk}.
    } %
%%%%%%%%%%%%
%%%%%%
\begin{align}
    \alpha_{\pm} &\approx  \frac{2\FF}{\pi} \lb( 2 k  \Delta \ell  \pm \gA \Delta \phi \rb)  . \label{eq:phaseShiftApprox}
\end{align}
%%%%%%

For two cavities both run near resonance and driven with oppositely handed light, a static phase difference $\Delta \alpha \equiv \alpha_{+} - \alpha_{-} = \Delta \alpha_\phi + \Delta \alpha_{\ell}$ develops, where
%%%%%%
\begin{align}
    \Delta \alpha_\phi &\equiv \frac{4\FF}{\pi} \cdot \gA \Delta \phi\: ,  \\
    \Delta \alpha_{\ell}  &\equiv \frac{4\FF}{\pi} \cdot k \lb( \Delta \ell_+ - \Delta \ell_- \rb) ,
\end{align}
%%%%%%
and where $\Delta \ell_\pm$ are the offsets from resonant length for the cavities driven with $\pm$-handed light.
The axion signal $\Delta \alpha_{\phi}$ is a static, achromatic phase difference between the cavities, while a differential length fluctuation of the cavities causes a chromatic phase difference $\Delta \alpha_{\ell} = \Delta \alpha_{\ell}(\lambda)$.

To estimate $\Delta \phi$, we adopt a simplified model for Earth (radius $R_{\oplus}$), ignoring its internal structure and approximating it as a homogeneous, spherically symmetric body with baryon and lepton charges satisfying $B\approx 2 L\approx M_{\oplus}/\mu_a$, where $M_{\oplus}$ is the mass of Earth and $\mu_a$ is the atomic mass unit.
In this setup, the on-resonance axion-induced phase difference is
%%%%%%
\begin{align}
    \Delta \alpha_{\phi}  &= \gEff\gA  \frac{\FF}{\pi^2} \frac{M_{\oplus}}{\mu_a} \frac{\ell \, e^{-m_\phi d}}{(R_{\oplus}+d)^2} \Psi(m_\phi \ell,m_\phi R_{\oplus},m_\phi d)\: ,\label{eq:DeltaAlphaExact}
\end{align}
%%%%%%
where $d$ is the distance between the surface of Earth and the lower mirrors of the FP cavities (we take $d = 10 \, \text{cm} \sim \lb[ 2\,\mu\text{eV} \rb]^{-1}$ throughout), $\gEff \equiv \gB + \frac{1}{2} (\gBL+\gL)$, and 
%%%%%%
\begin{align}
   \begin{split}
   \Psi(x,y,z) &\equiv - \frac{3e^{-y}}{y^3}  \lb( y \cosh y - \sinh y \rb) \\ &\qquad \times \frac{y+z}{x}  \lb[ 1 - \frac{e^{-x}}{1+x/(y+z)} \rb].\label{eq:PsiDefn}
   \end{split}
\end{align}
%%%%%%
$\Psi$ accounts for $\phi(r)$ being sourced dominantly by those parts of Earth within the axion-field range $r_{\phi}$~\cite{Adelberger:2003zx,Berge:2017ovy}, and for the full $r$-dependence of $\phi(r)$.
If $\ell,d \ll R_{\oplus}$ (the physical case), we have $\Delta \alpha_\phi \propto (m_\phi)^{-n}$ with $n=0$ for $m_{\phi} \lesssim 1/R_{\oplus}$, and $n=1$ for $1/R_{\oplus} \lesssim m_{\phi} \lesssim \min\lb[ 1/\ell , 1/d\rb]$. 
If $d<\ell$, then $n=2$ for $1/\ell \lesssim m_{\phi} \lesssim 1/d$.
For $m_\phi \gtrsim 1/d$, $\Delta \alpha_\phi \propto e^{-m_\phi d}$.

For Earth, and for $m_{\phi} \ell , m_{\phi}d \ll m_{\phi}R_{\oplus} \ll 1$, we have that
%%%%%%
\begin{align}
    \begin{split}
    \Delta \alpha_{\phi} &\sim  6\times 10^{-12} \times \lb( \frac{\FF}{10^4} \rb) \times \lb( \frac{\ell}{1\,\text{m}} \rb) \\ 
    &\quad \times \lb( \frac{ \gEff }{6\times 10^{-25}} \rb) \times \lb( \frac{ \gA }{6\times 10^{-13} \,\text{GeV}^{-1}} \rb)\: .
    \end{split}
\end{align}
%%%%%%

We envisage standard dark-port operation for the Michelson interferometer~\cite{Maggiore:2007zz}: carrier-signal phase modulation creates sidebands displaced from FP cavity resonance that do not pick up a phase difference, whereas the carrier is tuned to FP cavity resonance and experiences the signal phase difference.
Access to a linear signal is achieved in the beat note of the carrier and the sidebands at the beamsplitter, which can be mixed down for readout.

The signal in \eqref{eq:DeltaAlphaExact} is static, but if the orientation of the cavities is rotated from vertical about their midpoint at an angular frequency $\Omega$, the signal is modulated up to finite frequency: $\Delta \alpha_\phi \rightarrow \Delta \alpha_\phi \cdot f(t)$, where, for $\ell \ll R_{\oplus}$, 
%%%%%%
\begin{align}
    f(t) &
    \approx \frac{\sinh\lb[ \frac{m_\phi \ell}{2} \cos\lb( \Omega t \rb) \rb] }{\sinh\lb( \frac{m_\phi \ell}{2} \rb)}
    \stackrel{m_\phi \ell\ll1}{\longrightarrow} \cos\lb( \Omega t \rb).\label{eq:modulation}
\end{align}
%%%%%%
Were the cavities instead modulated at a frequency $\Omega$ by an angle $\pm \theta_0$ from vertical, $\Omega t \rightarrow \theta_0 \cos(\Omega t)$ in \eqref{eq:modulation}.

For sufficiently light axions ($m_\phi \lesssim 1/\text{AU}\sim 10^{-18}\,\eV$), one could additionally search for the sub-leading axion gradient sourced by the Sun at Earth, $|\nabla \phi|_{\odot} \sim 10^{-4}\,\eV^2 \times \gB/(6\times 10^{-25})$ for $m_\phi \lesssim 1/\text{AU}$, that modulates direction at the period of a synodic day.

One can also consider using a small, laboratory mass to source a (much smaller) axion field gradient. 
For cavities in radial orientation around a homogeneous spherical source of radius $R$ (with the nearest mirror of each FP cavity a distance $d$ from the source surface), the signal is given by \eqref[s]{eq:DeltaAlphaExact} and (\ref{eq:PsiDefn}) with the replacements $R_{\oplus} \rightarrow R$ and $M_\oplus \gEff / \mu_a \rightarrow  B \gB + L\gL + (B-L) \gBL$.
Other source geometries give similar results, up to geometrical factors.
Importantly, in this setup the cavities can be oriented horizontally, with modulation of the signal to non-zero frequency achieved either by moving the source mass (cf.~\citeR{Arvanitaki:2014dfa}) or by rotating the cavities about a vertical axis (cf.~\citeR{Wagner:2012ui}).
This may have technical advantages over vertically oriented cavities being rotated about a horizontal axis to achieve modulation of the signal from the fixed, vertical Earth-sourced field gradient.

%%%%%%%%%%%%%%%%%%%%%%%%%%%%%%%%%%%%%%%%%%%%%%%%%%%%%%%%%%%%%%%%%%%%%%%%%%%%%%%%%%%%%%%%%%
%%%%%%%%%%%%%%%%%%%%%%%%%%%%%%%%%%%%%%%%%%%%%%%%%%%%%%%%%%%%%%%%%%%%%%%%%%%%%%%%%%%%%%%%%%
\section{Reach}%
\label{sec:reach}%
%%%%%%%%%%%%%%%%%%%%%%%%%%%%%%%%%%%%%%%%%%%%%%%%%%%%%%%%%%%%%%%%%%%%%%%%%%%%%%%%%%%%%%%%%%
%%%%%%%%%%%%%%%%%%%%%%%%%%%%%%%%%%%%%%%%%%%%%%%%%%%%%%%%%%%%%%%%%%%%%%%%%%%%%%%%%%%%%%%%%%
A projection for the reach of this experiment to the axion signal requires understanding various stochastic noise sources and confounding systematics, and in some cases mitigating the latter.

Photon shot noise statistically limits interferometer sensitivity.
Suppose we integrate coherently for a time~$\tau$; the expected number of photons arriving at the interferometer beamsplitter, where the interference pattern is observed, is $N_{\gamma} = P_0 \tau / \omega$, where $P_0 \sim \pi P_{\text{cav}}/(2\FF)$ is the average power incident on the beamsplitter if $P_{\text{cav}}$ is the circulating FP cavity power and $\omega = 2\pi / \lambda$ is the angular frequency of the light.
The phase uncertainty is then~$\sim 1/\sqrt{N_{\gamma}}$. 
We estimate that the signal-to-noise ratio (SNR) for shot noise is
%%%%%%
\begin{align}
     \text{SNR}_{\text{shot}} \sim \sqrt{\frac{2\pi P_{\text{cav}} \tau}{\omega\FF}} \Delta \alpha_\phi \: .\label{eq:ShotSNR}
\end{align}
%%%%%%
For all of our projections, we take $\tau = 300 \, \text{days}$ and fix $P_\text{cav} = 1\, \text{MW}$, adjusting $P_0$ depending on $\FF$.

In distinct contrast to an architecture optimized for gravitational-wave detection, the mirrors at the ends of the FP cavities in our proposal should be attached rigidly to a structure instead of being isolated; a high mechanical-resonance frequency for this support structure will suppress radiation-pressure noise. 
Modeling the cavity system as being embedded in a rigid block of mass $M$ with fundamental vibrational frequency $\omega_{\text{vib}}$, we can estimate the SNR due to radiation-pressure-induced length fluctuations as 
%%%%%%
\begin{align}
    \text{SNR}_{\text{rad}} &\sim \frac{\pi}{2\sqrt{2}} \frac{M \omega_{\text{vib}}^2}{\FF^{3/2}} \sqrt{\frac{\tau}{ P_{\text{cav}} \omega^3}} \Delta \alpha_\phi\: .
\end{align}
%%%%%%
We take the block to be cubic, with side-length $\ell_{\text{sys}}$, constant density $\rho_{\text{cav}} \sim 8 \, \text{g}/ \text{cm}^3$, and sound speed $c_s \sim 6\, {\text{km}} / {\text{s}}$, and we estimate $\omega_{\text{vib}} \sim \pi {c_s} / {\ell_\text{sys}}$.

For a cavity system at finite temperature there will also be thermal vibrational noise leading to fluctuating cavity lengths. 
Taking the same cavity-system model as above, we use the fluctuation--dissipation theorem to estimate 
%%%%%%
\begin{align} 
    \text{SNR}_{\text{vib}} \sim \frac{\pi}{4} \sqrt{\frac{\tau M Q_\text{vib} \omega_{\text{vib}}^3}{T_{\text{sys}}}} \frac{\Delta \alpha_\phi}{\omega \FF}\: , \label{eq:thermVib}
\end{align}
%%%%%%
where $T_{\text{sys}}$ is the system temperature and $Q_\text{vib}$ is the mechanical-resonance quality factor. 
We assume cryogenic operation,%
%%%%%%%%%%%%
\footnote{\label{ftnt:RFtemp}%
    We note that some of the RF systems we discuss later may need to have the entire experiment cooled to $\lesssim 1.4\,$K in order to attain ultra-high finesse.
    Standard cooling techniques are able to achieve such temperatures; see, e.g., \citeR{Romanenko:2023irv}.
    We nevertheless give a more conservative thermal vibrational noise estimate based on a higher, $4$\,K temperature.
    Due to the $\propto 1/\sqrt{T_{\text{sys}}}$ scaling in \eqref{eq:thermVib}, lowering $T_{\text{sys}}$ from 4\,K to $\sim 1.4\,$K would only improve the thermal vibrational SNR by a factor of $\sim 2$; as we advance only benchmarks here, this is well within the margins to which our projections should be understood to be certain.
} %
%%%%%%%%%%%%
$T_{\text{sys}} \sim 4 \, \text{K}$,
and take $Q_{\text{vib}} \sim 10^3$.
Separately, we estimated the thermal readout noise for an RF system  via the Dicke radiometer equation and found it to be subdominant.

Differential cavity length fluctuations arising from cavity temperature fluctuations may be a serious limitation to this experimental approach.
It is however challenging to estimate the size of this effect without a more detailed technical experimental design; we therefore do not give a quantitative estimate here.
If both cavities are embedded in the same rigid structure, some common length-fluctuation noise could cancel (see also \citeR{DeRocco:2018jwe}).
A more massive cavity system also assists in providing greater thermal inertia.
We also discuss below a co-metrology technique that in principle allows the measurement and subtraction of differential length-fluctuation noise, leaving only shot noise to contend with.

Thermal mirror coating noise may be of relevance, especially for an optical setup.
However, this noise depends in some detail on the manner in which the mirrors are designed (see, e.g., \citeR{PhysRevD.81.122001}), the choice of an optical or RF system, and whether (and at what frequency) signal modulation can be achieved.
We thus defer estimates of the impact of such noise on this experiment to future technical work.

%%%%%%%%%%%%%%%%%%%%%%%%%%%%%%%%%%%%%%%%%%%%%%%%%%%%%%%%%%%%%%%%%%%%%%%%%%%%%%%%%%%%%%%%%%%%%
\begin{table}[t]
%%%%%%%%%%%%%%%%%%%%%%%
\caption{\label{tab:Parameters}
    %%%%%%%%%%%%%%%%%%%%%%%%%%%%%%%%%%
    Experimental parameters, as defined in the text, assumed for the optical or open RF Fabry--P{\'e}rot interferometer reach projections. 
    See the discussion in the text regarding the choices of parameters for the RF systems.%
    %%%%%%%%%%%%%%%%%%%%%%%%%%%%%%%%%%
    }
%%%%%%%%%%%%%%%%%%%%%%%%%%%%%%%%%%%%%%%%%%%%%%%%%%%%%%%%%%%%%%%%%%%%%%%%%%%%%%%%%%%%%%%%%%%%
\begin{ruledtabular}
\begin{tabular}{llll}
%%%%%%%%%%%%%%%%%%%%%%%%%%%%%%%%%%%%%%%%%%%%%%%%%%%%%%%%%%%%%%%%%%%%%%%%%%%%%%%%%%%%%%%%%%%%
Parameter               &   Optical             &    Conservative RF    &   Optimistic RF                 \\ \hline\hline
%%%%%%%%%%%%%%%%%%%%%%%%%%%%%%%%%%%%%%%%%%%%%%%%%%%%%%%%%%%%%%%%%%%%%%%%%%%%%%%%%%%%%%%%%%%%
$\omega/2\pi$           &   1/(1064\,nm)        &   51\,GHz             &   51\,GHz            \\
$\FF$                   &   $10^4$              &   $10^7$              &   $4.6\times 10^9$   \\
$\ell$ [m]              &   1                   &   $0.027$             &   $0.027$            \\
$\ell_{\text{sys}}$ [m] &   1                   &   $0.3$               &   $0.3$                                  
%%%%%%%%%%%%%%%%%%%%%%%%%%%%%%%%%%%%%%%%%%%%%%%%%%%%%%%%%%%%%%%%%%%%%%%%%%%%%%%%%%%%%%%%%%%%
\end{tabular}
\end{ruledtabular}
\end{table}
%%%%%%%%%%%%%%%%%%%%%%%%%%%%%%%%%%%%%%%%%%%%%%%%%%%%%%%%%%%%%%%%%%%%%%%%%%%%%%%%%%%%%%%%%%%%

%%%%%%%%%%%%%%%%%%%%%%%%%%%%%%%%%%%%%%%%%%%%%%%%%%%%%%%%%%%%%%%%%%%%%%%%%%%%%%%%%%%%%%%%%%
\begin{figure*}[t]
    \centering
    \includegraphics[width=0.47\textwidth]{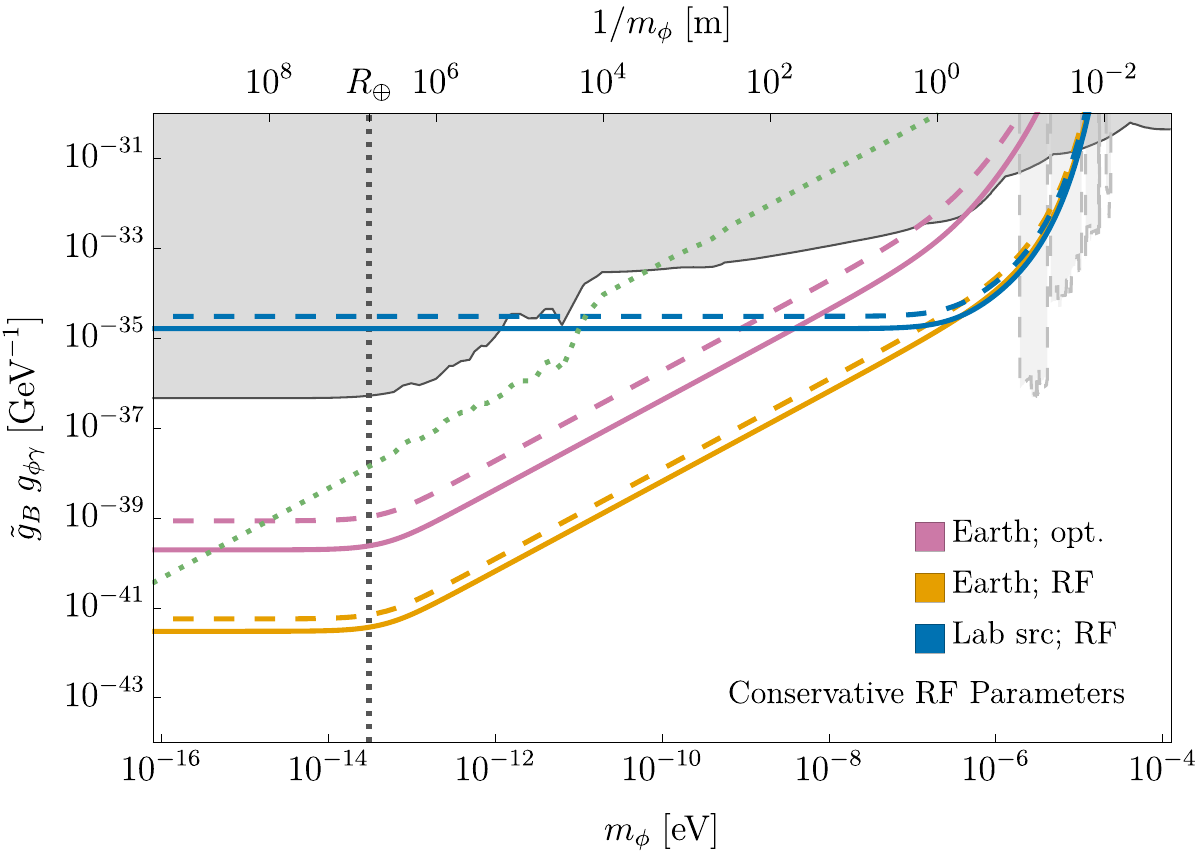}
    \includegraphics[width=0.47\textwidth]{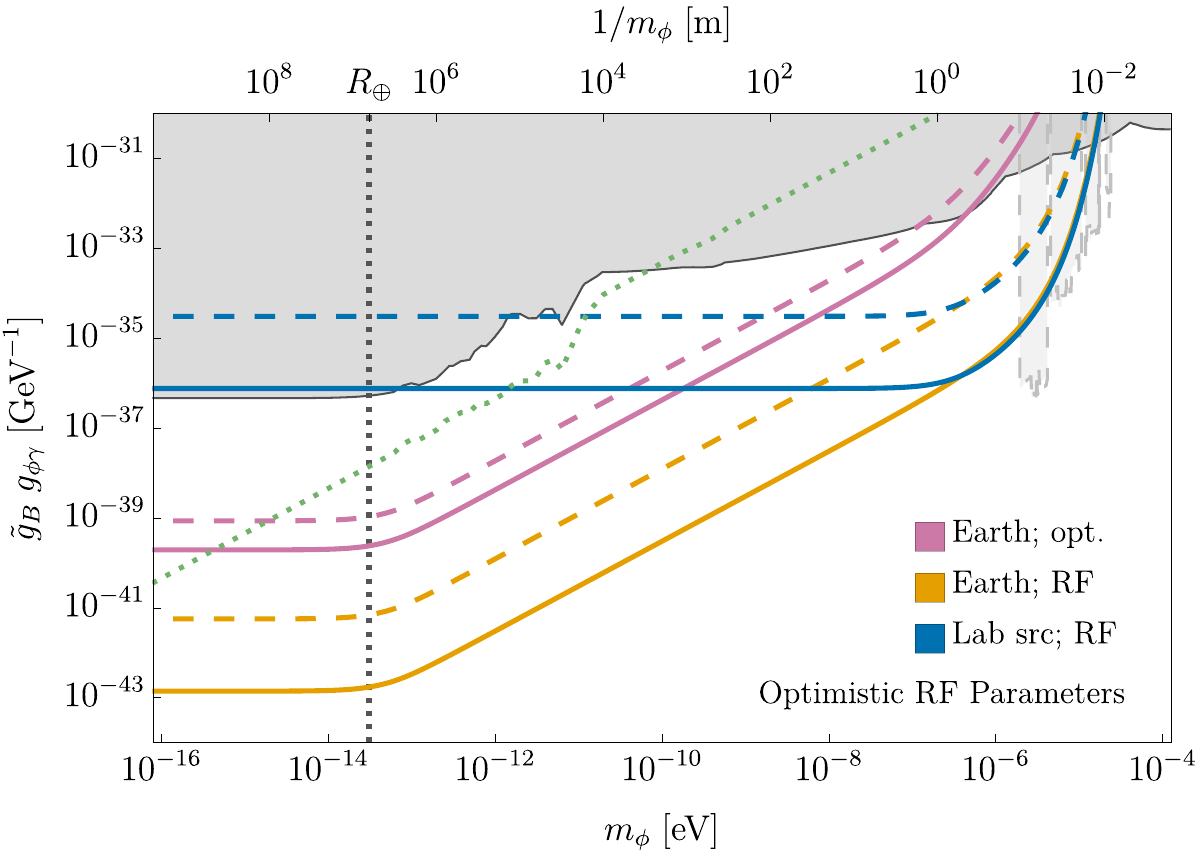}
    \caption{\label{fig:sensitivity}%
        %%%%%%%%%%%%%%%%%%%%%%%%%%%%%%%%%%
        Projected reach curves at an SNR of 1 for various experimental architectures: 
        Earth as a source and a laser-based cavity search (purple); 
        Earth as a source and an RF-based cavity search (orange); and 
        a laboratory source and an RF-based cavity search (blue).
        Parameters assumed are discussed in the text; see also \tabref{tab:Parameters}.
        In the left panel, we use the conservative parameters for the RF system from \tabref{tab:Parameters}; in the right panel, we use the optimistic parameters.
        Dashed lines are limited by vibrational noise at finite temperature [\eqref{eq:thermVib}], while solid lines take into account only shot noise [\eqref{eq:ShotSNR}].
        The scaling of all projections with $m_\phi$ arises from the $\Delta \alpha(m_\phi)$ behavior explained below \eqref{eq:PsiDefn}; the sharp high-mass cutoff occurs for $m_{\phi} \gtrsim d^{-1}$.
        Independent limits arise on $\gB$ from fifth-force experiments~\cite{Schlamminger:2007ht,ADELBERGER2009102,Wagner:2012ui,PhysRevLett.129.121102,Touboul:2017grn,Berge:2017ovy,Shaw:2021gnp,Fedderke:2022ptm}, and on $\gA$ from various astrophysical constraints~\cite{Wouters:2013hua,Marsh:2017yvc,Reynolds:2019uqt,Reynes:2021bpe,Dessert:2022yqq,Jaeckel:2017tud,Hoof:2022xbe,Payez:2014xsa,Ayala:2014pea,Dolan:2022kul}; taking a simple product of these constraints rules out the dark-grey region.
        In the same fashion, the light-grey shaded region would be excluded by haloscope constraints~\cite{RBFUF,ADMX:2009iij,ADMX:2018gho,ADMX:2018ogs,ADMX:2019uok,ADMX:2021nhd,ADMX:2021mio} were the same axion species also all of the DM.
        Above the dotted green line, an axion mass tuning may be required, assuming $\Lambda = 10$\,TeV and $\gA$ fixed at current limits.
        %%%%%%%%%%%%%%%%%%%%%%%%%%%%%%%%%%
        }
\end{figure*}
%%%%%%%%%%%%%%%%%%%%%%%%%%%%%%%%%%%%%%%%%%%%%%%%%%%%%%%%%%%%%%%%%%%%%%%%%%%%%%%%%%%%%%%%%%

The total SNR may be estimated as
%%%%%%
\begin{align}
    \frac{1}{\lb(\text{SNR}_\text{tot}\rb)^{2}} = \sum\limits_{i} \frac{1}{\lb( \text{SNR}_i \rb)^{2}}.
\end{align}
%%%%%%
In \figref{fig:sensitivity}, we show $\text{SNR}_{\text{tot}} = 1$ (dashed lines) and  $\text{SNR}_{\text{shot}} = 1$ (solid lines) reach estimates for the product of couplings $\gB \gA$ for three different experimental architectures, using either the optical or open RF cavity interferometer parameters (either conservative or optimistic; see discussion below) shown in \tabref{tab:Parameters}:
(1) Earth as the axion source, with the optical interferometer setup;
(2) Earth as the axion source, with the two RF interferometer setups; and
(3) a small laboratory-mass axion source (radius $R = 1 \, \text{m}$, density $\rho = 8 \,\text{g}/\text{cm}^3$), with the same two RF interferometer setups.

In making our reach projections, we set $\gL = \gBL = 0$; current constraints and signal-reach estimates for cases with these couplings non-zero are broadly similar to those for $\gB$, up to $\mathcal{O}(3)$ numerical factors.
Note that the lab-sourced axion gradient in case~(3) is far weaker than that for Earth, $|\nabla\phi|_{\text{src}}(r=R) \sim 2\times 10^{-7} |\nabla \phi|_{\oplus}$ for $m_{\phi}R_{\oplus} \lesssim 1$; it is also significantly weaker than the naturally modulating solar source, but the latter is only detectable for $m_\phi \lesssim 10^{-18} \,\eV$.
For a homogeneous source, $\Delta \alpha_{\phi} \propto \rho R$ for $m_{\phi} R\ll 1$; tradespace thus exists to optimize for lab-source size and/or density.

Some comments are in order regarding our choices for benchmarks for the RF system.
We adopt both an `optimistic' and a `conservative' set of parameters; see \tabref{tab:Parameters}.
Our choices are informed by \citeR{KuhrUltrahighFinesse}.
For both benchmarks, we adopt the $\ell$ and $\omega$ parameters used for the ultra-high-finesse cavities demonstrated in \citeR{KuhrUltrahighFinesse}.
For the `optimistic' setup, additionally adopt the same ultra-high $\FF$ demonstrated in \citeR{KuhrUltrahighFinesse}.

In order for an FP cavity to support circular polarization states, its two orthogonal linear polarization modes must be simultaneously resonant within their linewidth.
The actual cavity in \citeR{KuhrUltrahighFinesse} was intentionally designed (for unrelated reasons) with its orthogonal linear polarization states split in frequency by $\sim 1.2\,\text{MHz}$, which is much larger than the $\sim 3\,$Hz mode linewidths; that specific cavity is thus not able to support circular polarization modes.
However, this splitting was achieved in \citeR{KuhrUltrahighFinesse} by intentionally machining the mirrors to have a toroidal shape with radii of curvature on orthogonal axes offset by $\pm 0.6$\,mm from 40\,mm.
But this mirror was formed to this shape with a $< 300\,$nm peak--valley shape accuracy, and 10\,nm local surface roughness.
Indeed, even much larger, meter-scale optical mirrors can be machined to similar $\sim 330$\,nm peak--valley shape accuracy (with rms shape accuracy around 15\,nm after reflective coating, and local surface roughness at sub-nm levels)~\cite{Zhang:2022lkb}.
It is thus entirely plausible that mirrors with a finesse similar to those of \citeR{KuhrUltrahighFinesse} could be formed exhibiting a much higher degree of axial symmetry than intentionally selected in \citeR{KuhrUltrahighFinesse}.
Assuming that the frequency splitting is linear in the difference between the largest and smallest radii of curvature of the mirror (at least for small such differences), a na\"ive estimate then suggests that the orthogonal polarization modes could plausibly be brought to within $\mathcal{O}(10^2\,\text{Hz})$ via machining precision alone.
It may then be possible to bring the two linearly polarized cavity modes into simultaneous resonance via, e.g., piezoelectric actuation to deform the mirrors.%
%%%%%%%%%%%%
\footnote{\label{ftnt:tuningHz}%
    While the deformation of mirror shaping to bring two modes into simultaneous resonance is naturally a more complicated manipulation, open (closed) cavity tuning control for a single mode at the few Hz level or better via piezoelectric actuation is shown in \citeR{KuhrUltrahighFinesse} (\citeR{Romanenko:2023irv}); cavity frequency drifts over time at the level of a few Hz per hour are however also observed in both cases, and short-period frequency jitters also at the few Hz level are reported in \citeR{Romanenko:2023irv} for a closed resonator architecture.
    } %
%%%%%%%%%%%%
However, we acknowledge that stable control of cavity modes at such high finesse may be a technical challenge.

On the other hand, were such a cavity run at lower finesse, linewidths would broaden and it would become easier to bring the two orthogonal linear polarization modes into simultaneous resonance, allowing support for the necessary circular modes.
In our `conservative' benchmark, we thus adopt a finesse a factor of $\sim 460$ smaller than that demonstrated in \citeR{KuhrUltrahighFinesse}; this broadens the cavity linewidths to $\mathcal{O}(\text{kHz})$.
On the basis of the discussion above, arranging the linear modes to overlap within this linewidth would likely be significantly easier.
Achieving lower finesse would not necessarily require a new cavity or mirrors: e.g., simply increasing the temperatures of the RF mirrors used in \citeR{KuhrUltrahighFinesse} to roughly 3\,K would appear to suffice%
%%%%%%%%%%%%
\footnote{\label{ftnt:thankReferee}%
    We thank an anonymous referee for pointing this out.} %
%%%%%%%%%%%%
(the RF mirrors in \citeR{KuhrUltrahighFinesse} attained a low-temperature plateau finesse value below $\sim 1.4$\,K; above this, $\FF$ fell exponentially with increasing temperature).
We defer in-depth consideration and resolution of these issues to future work, and here merely present our conservative and optimistic parameter selections as benchmarks.

Finally on this point, we note that even with the more conservative RF assumptions, the shot-noise limited sensitivity for the RF cavities is still mildly dominated by the thermal vibrational noise estimate, although (as discussed in the next section) the latter may possibly be mitigated by co-metrology techniques.

Because the scalar coupling breaks the axion shift symmetry, \figref{fig:sensitivity} also shows where the axion bare mass may need to be tuned against the resulting quantum correction $\delta m^2 \sim \gB^2 \Lambda^2 / (8\pi^2)$, where $\Lambda$ is a UV cutoff scale, to maintain a light axion, assuming that $\gA$ is fixed at current bounds.

%%%%%%%%%%%%%%%%%%%%%%%%%%%%%%%%%%%%%%%%%%%%%%%%%%%%%%%%%%%%%%%%%%%%%%%%%%%%%%%%%%%%%%%%%%
%%%%%%%%%%%%%%%%%%%%%%%%%%%%%%%%%%%%%%%%%%%%%%%%%%%%%%%%%%%%%%%%%%%%%%%%%%%%%%%%%%%%%%%%%%
\section{Discussion}%
\label{sec:discussion}%
%%%%%%%%%%%%%%%%%%%%%%%%%%%%%%%%%%%%%%%%%%%%%%%%%%%%%%%%%%%%%%%%%%%%%%%%%%%%%%%%%%%%%%%%%%
%%%%%%%%%%%%%%%%%%%%%%%%%%%%%%%%%%%%%%%%%%%%%%%%%%%%%%%%%%%%%%%%%%%%%%%%%%%%%%%%%%%%%%%%%%
The achromatic axion phase difference can be distinguished from chromatic noise arising from differential cavity-length fluctuations.
Running each cavity with multiple cavity-resonant frequencies of light simultaneously could thus suppress chromatic backgrounds; this differs from a GW detector, where the signal is degenerate with a differential length fluctuation.
Such multi-frequency co-metrology techniques (cf., e.g., \citeR[s]{Safronova:2017xyt,Terrano:2021zyh}) could, in principle, allow the mitigation of differential length-fluctuation noise backgrounds to the level of the shot-noise floor.
In particular, length fluctuations due to thermal expansion may be severe enough to require such an approach.
Further investigation of this technique is therefore warranted.

We also note that there is a further potential systematic for the measurement: cavities have a degree of intrinsic linear birefringence; see, e.g., \citeR[s]{PhysRevA.62.013815,PhysRevA.93.013833,Ejlli:2017lwm,Ejlli:2020yhk}.
This strongly motivates signal modulation: only variations in the cavity birefringence properties at the modulation frequency $\Omega$ are then relevant (including possible changes in stress-induced birefringence were the cavities to be rotated in Earth's gravitational field).
Moreover, the multi-frequency techniques noted above would also assist in breaking degeneracy with the signal, as intrinsic birefringence will be chromatic.
Additionally, running the cavity at multiple different lengths $\ell$ could break degeneracy between our signal $\Delta \alpha_\phi \propto \ell$, and any $\ell$-independent phase difference due to mirror coatings or abnormalities.
However, because cavity birefringence depends in detail on the properties of the optical or RF elements in the cavity, we defer study of this issue to future technical work.

In addition to the open RF FP cavities we considered in this work, another possible experimental architecture would use closed superconducting radio-frequency (SRF) cavity resonators (see, e.g., \citeR[s]{Padamsee:2014bfa,Romanenko:2023irv}) in the interferometer arms.
However, finite machining tolerances of the cavity walls appear to limit the fidelity with which opposite circular polarizations propagate without mixing, limiting the usable cavity finesse to well below that attainable for the highest-$Q$ SRF cavities [$\FF = (\omega_{\textsc{fsr}}/\omega_n)Q$, where $Q$ is the cavity quality factor, $\omega_{\textsc{fsr}}$ is the cavity free spectral range, and $\omega_n$ is the cavity resonance frequency].

In this paper, we have proposed a novel interferometric experiment to search for static axion field gradients.
Provided that shot-noise-limited sensitivity can be reached, an approach exploiting open, high-finesse RF FP resonators could allow the exploration of $\sim 5$ (or more) orders of magnitude of new parameter space beyond current limits on the product coupling $\gQ \gA$ for $m_\phi \lesssim 4\times 10^{-11} \, \eV$.\\

%%%%%%%%%%%%%%%%%%%%%%%%%%%%%%%%%%%%%%%%%%%%%%%%%%%%%%%%%%%%%%%%%%%%%%%%%%%%%%%%%%%%%%%%%%
%%%%%%%%%%%%%%%%%%%%%%%%%%%%%%%%%%%%%%%%%%%%%%%%%%%%%%%%%%%%%%%%%%%%%%%%%%%%%%%%%%%%%%%%%%
\acknowledgments%
%%%%%%%%%%%%%%%%%%%%%%%%%%%%%%%%%%%%%%%%%%%%%%%%%%%%%%%%%%%%%%%%%%%%%%%%%%%%%%%%%%%%%%%%%%
%%%%%%%%%%%%%%%%%%%%%%%%%%%%%%%%%%%%%%%%%%%%%%%%%%%%%%%%%%%%%%%%%%%%%%%%%%%%%%%%%%%%%%%%%%
We thank William DeRocco, Anson Hook, Xuheng Luo, Anubhav Mathur, and Oleksandr S.~Melnychuk for useful discussions.
We also thank an anonymous referee for constructive comments.
This work was supported by the U.S.~Department of Energy~(DOE), Office of Science, National Quantum Information Science Research Centers, Superconducting Quantum Materials and Systems Center~(SQMS) under Contract No.~DE-AC02-07CH11359. 
Fermilab is operated by the Fermi Research Alliance, LLC under Contract No.~DE-AC02-07CH11359 with the~DOE.
D.E.K.\ and S.R.\ are supported in part by the U.S.~National Science Foundation~(NSF) under Grant No.~PHY-1818899.
S.R.\ is also supported by the~DOE under a QuantISED grant for MAGIS.
The work of S.R.\ and M.A.F.\ was also supported by the Simons Investigator Award No.~827042. 
Research at Perimeter Institute is supported in part by the Government of Canada through the Department of Innovation, Science and Economic Development and by the Province of Ontario through the Ministry of Colleges and Universities.
M.A.F.\ gratefully acknowledges the hospitality of the Simons Center for Geometry and Physics at Stony Brook University, where part of this work was undertaken.
J.O.T.\ would like to thank the Kavli Institute for the Physics and Mathematics of the Universe for hospitality during the completion of this work.

%%%%%%%%%%%%%%%%%%%%%%%%%%%%%%%%%%%%%%%%%%%%%%%%%%%%%%%%%%%%%%%%%%%%%%%%%%%%%%%%%%%%%%%%%%
%%%%%%%%%%%%%%%%%%%%%%%%%%%%%%%%%%%%%%%%%%%%%%%%%%%%%%%%%%%%%%%%%%%%%%%%%%%%%%%%%%%%%%%%%%
\bibliography{references.bib}
%%%%%%%%%%%%%%%%%%%%%%%%%%%%%%%%%%%%%%%%%%%%%%%%%%%%%%%%%%%%%%%%%%%%%%%%%%%%%%%%%%%%%%%%%%
%%%%%%%%%%%%%%%%%%%%%%%%%%%%%%%%%%%%%%%%%%%%%%%%%%%%%%%%%%%%%%%%%%%%%%%%%%%%%%%%%%%%%%%%%%

%%%%%%%%%%%%%%%%%%%%%%%%%%%%%%%%%%%%%%%%%%%%%%%%%%%%%%%%%%%%%%%%%%%%%%%%%%%%%%%%%%%%%%%%%%
%%%%%%%%%%%%%%%%%%%%%%%%%%%%%%%%%%%%%%%%%%%%%%%%%%%%%%%%%%%%%%%%%%%%%%%%%%%%%%%%%%%%%%%%%%
%%%%%%%%%%%%%%%%%%%%%%%%%%%%%%%%%%%%%%%%%%%%%%%%%%%%%%%%%%%%%%%%%%%%%%%%%%%%%%%%%%%%%%%%%%
%%%%%%%%%%%%%%%%%%%%%%%%%%%%%%%%%%%%%%%%%%%%%%%%%%%%%%%%%%%%%%%%%%%%%%%%%%%%%%%%%%%%%%%%%%
\end{document}